\documentclass[prl,aps,twocolumn,groupedaddress,showpacs,floatfix,preprintnumbers,superscriptaddress]{revtex4-1}
\usepackage{amsmath,amssymb,multirow,bm}

\newcommand{\beq}{\begin{equation}}
\newcommand{\eeq}{\end{equation}}
\newcommand{\bea}{\begin{eqnarray}}
\newcommand{\eea}{\end{eqnarray}}
\usepackage{graphicx}
\begin{document}
\preprint{}

\title{\bf Stellar oscillations induced by the passage of a fast stellar object}

\author{C.A. Bertulani,  M. Naizer, and W. Newton\\
\it Department of Physics and Astronomy, Texas A \& M University - Commerce, Commerce, TX 75429, USA}
\begin{abstract}
We investigate induced oscillations by the gravitational field of a fast stellar object, such as a neutron star or a black-hole in a near miss collision with another star. Non-adiabatic collision conditions may lead to large amplitude oscillations in the star. We show that for a solar-type star a resonant condition can be achieved by a fast moving stellar object  with velocity in the range of 100 km/s to 1000 km/s, passing at a distance of a few multiples of the star radius. Although such collisions are  rare, they are more frequent than head-on collisions, and their effects could be observed through a visible change of the star luminosity occurring within a few hours. 
\end{abstract}

\date{\today}

 
 \maketitle

Stellar oscillations, commonly known as pulsations, are understood in terms of modulation geared by the interaction of radiation with matter on its way from the center of the star \cite{Col03}.  Little is known about how other kinds of oscillations can be generated in stellar collisions, although an effort has been made in studying tidal oscillations due to the gravitational interaction with a companion in a binary system (see, e.g., \cite{ST83,Ter98,SW02}). Stellar collisions are often investigated in the context  of gravitational waves which could, in principle, be  detected by ground and space-based laser interferometers \cite{Abb08}.  Gamma-ray bursts arising from tidal disruption of neutron stars (NS) in NS-NS or NS-Black Hole (BH) collisions have also attracted interest \cite{LS74,Pac86,Goo86,Rob11,MB12}. 

The kinetic energy of a star with $m_\odot$ and velocity $v=1000$ km/s is about $10^{49}$ ergs, which is enough energy to power the luminosity of the Sun for one billion years. A central collision between such objects would be nothing less than spectacular. But if only a small fraction of this energy is transformed into the internal energy of a star during a relatively short time, the consequences would also be dramatic. This can be achieved in near miss collisions, with impact parameters larger than the sum of the radii of the collisional partners. There is plenty of available space for near misses, but very little room for central collisions and far collisions are much more frequent. It is therefore relevant to find out what are their consequences. Such processes have been previously studied for the specific purpose of assessing tidal oscillations in  NS \cite{ST83,Tsa13}. Here we show that a resonant condition arises for solar-like stars within the range of possible velocities, leading to effects within our observational reach.

\begin{figure}[tbp]
\begin{center}
\includegraphics[scale=0.42]{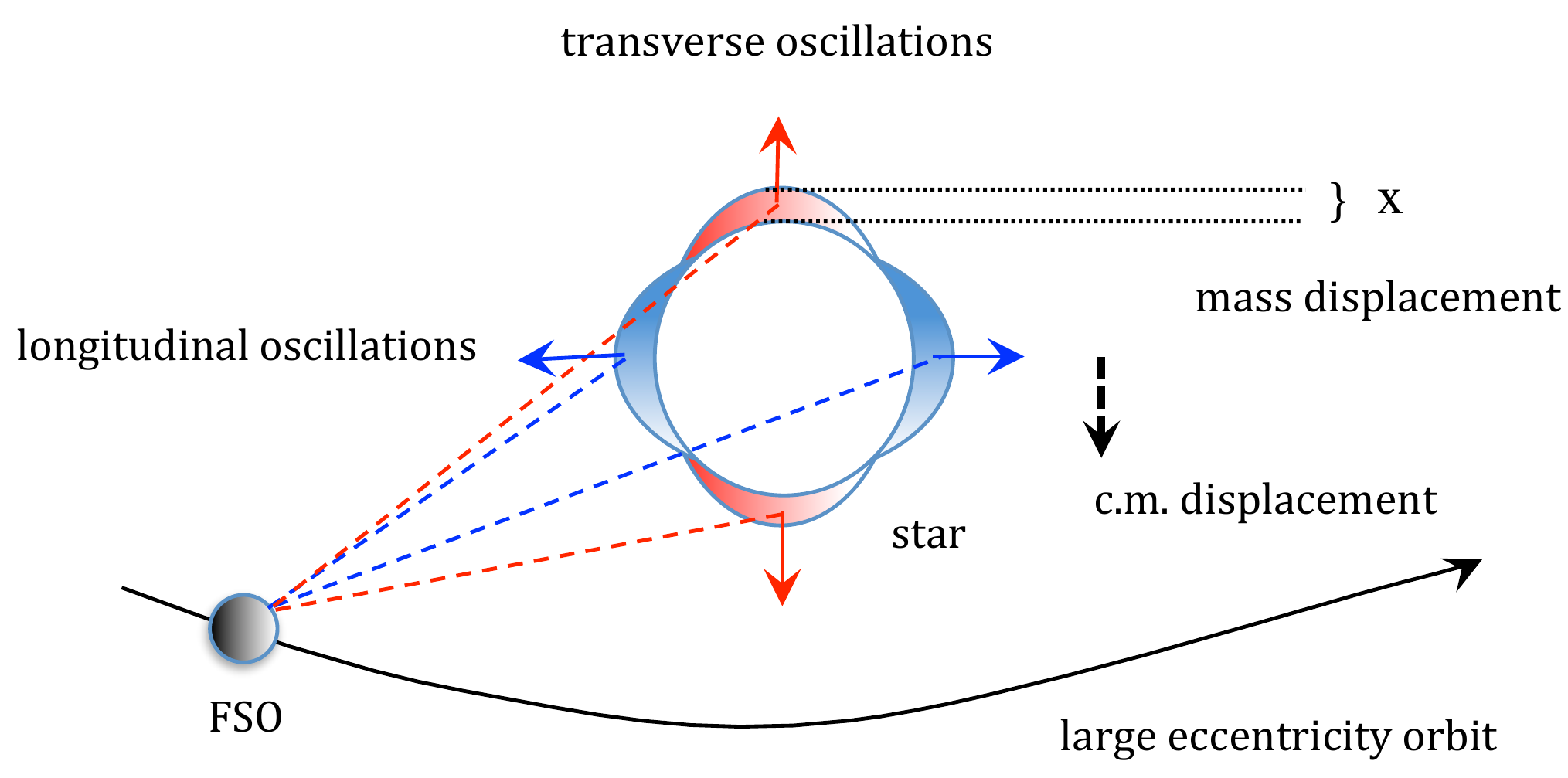}
\caption{Tidal oscillations induced in a star by the passage of a fast stellar object.} \label{fig:tosc}
\end{center}
\end{figure}

We consider the internal response of  a star due to the passage of a fast stellar object (FSO), e.g., a NS or a BH, at large impact parameters. This response can be modeled by considering the tidal force on a mass element  $dm_s$ of the star, roughly given by \cite{Avs77,Brad08}
$%\begin{equation}
d{F}\sim 2GM dm_s  x/ R^3, 
%\label{dF}
$ %\end{equation} 
where $M$ is the mass of the FSO, $G$ the gravitational constant, $R$ is the  distance between the center of mass of the FSO and the star, and $x$ is the  distance of the mass element $dm_s$ from the center of the star. The tidal force is best described by expanding the gravitational field of the FSO into multipoles, yielding at a position ${\bf x}$ inside the star \cite{Ja98}
\begin{equation}
V({\bf x},t)=-GM\sum_{lm}{4\pi \over 2l+1} Y_{lm}(\hat{\bf R}(t)) {x^l\over R^{l+1}(t)}Y_{lm}^*(\hat{\bf x}) . \label{Fvib4}
\end{equation}
In the center of mass of the star, the force on a mass element at ${\bf x}$ is obtained from the derivative of Eq. \eqref{Fvib4} with respect to $x$. The distance $R(t)$ is a function of time, and $t=0$ is  taken when the two stars are at the periapsis, or distance of closest approach. In the frame of reference of the star, the tidal force  acts in opposite sides from its center, trying to elongate it and leading to a  time-dependent ellipsoidal shaped oscillation. To lowest order, the passage of a FSO will induce quadrupole shaped vibrations, as seen in Figure \ref{fig:tosc}. Higher multipole vibrations such as octupole oscillations are also possible, but are orders of magnitude smaller and have been neglected here.

The time dependence of $R(t)$ and $\theta(t)$ is described by a hyperbolic Kepler trajectory, parametrized by
an orbital eccentricity $\epsilon >1$,  where $\theta(t)$ is the angular position of the FSO measured from the center of mass of the system and with respect to the line joining  it to the star so that, at $t=0$, $\theta = 0$, and $R=a$, the distance at the periapsis. For a collision with impact parameter $b$ one has $a=2b/(\alpha +\sqrt{\alpha^2 +4})$ where $\alpha=GMm_s/(Eb)$, $E=\mu v^2/2$ is the collision energy and $\mu=Mm_s/(m_s+M)$ is the reduced mass. The relation between the angular position and time can be obtained  solving coupled equations for $R$  and $t$ along the trajectory (see, e.g., Ref \cite{Mou84}). 

\begin{figure}[tbp]
\begin{center}
\includegraphics[scale=0.24]{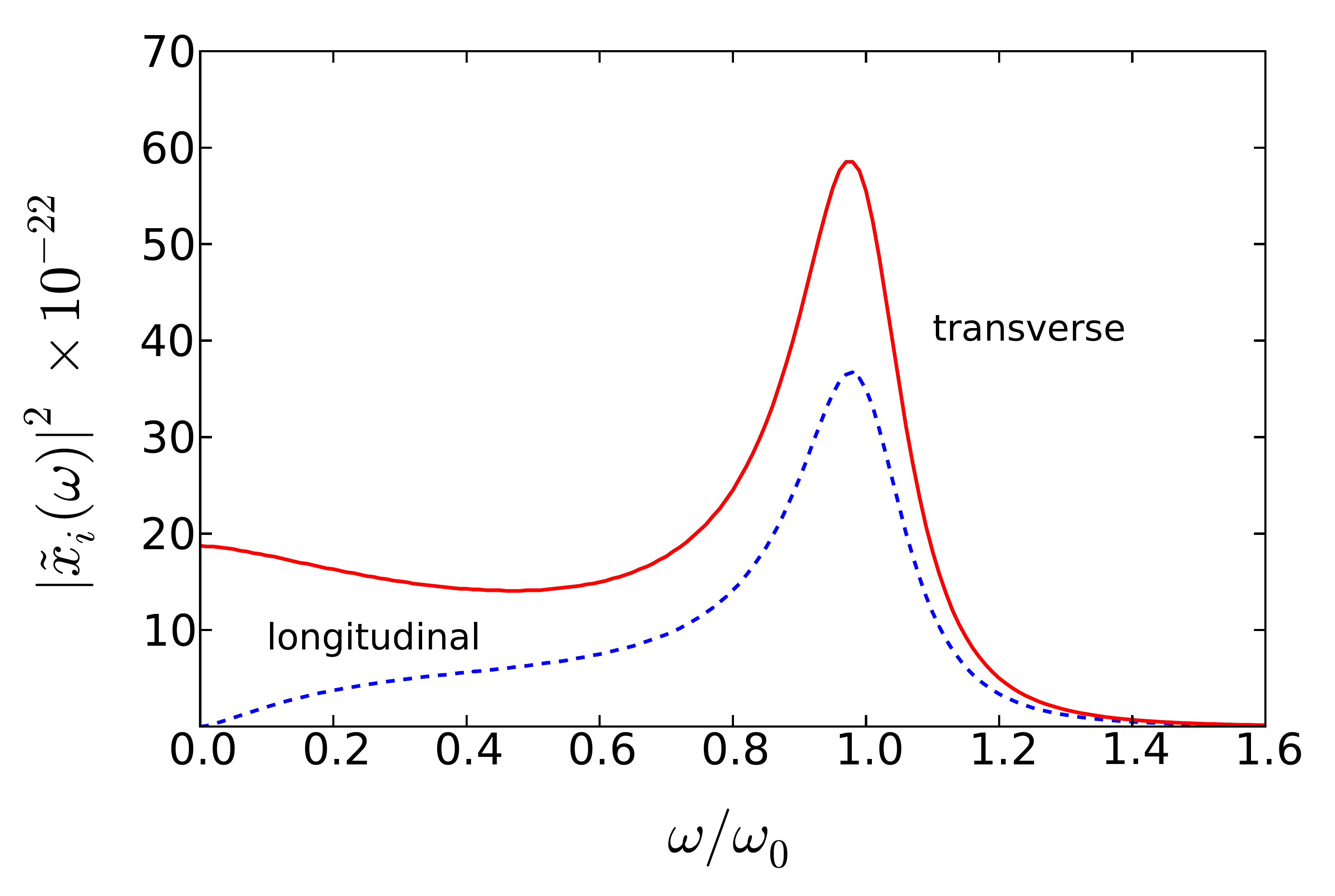}
\caption{The amplitudes $|\tilde{x}_i(\omega)|^2$ multiplied by $10^{-22}$ (m$^2$s$^2$ units) as a function of the ratio of the frequency $\omega$ and the natural oscillation frequency $\omega_0$, for longitudinal (dashed line) and transverse (solid line) oscillations.} \label{fig:xi}
\end{center}
\end{figure}

Non-radial stellar oscillation modes  can be described with hydrodynamical models to high accuracy (see, e.g., \cite{Dals14}). We have no reason to use such models, as the physical situation we consider here has never been observed before. We adopt a simple model including only (a) a single inertia parameter, (b) a linear restoring force, and (c) a damping parameter. The simplest model of this kind was developed by Lord Kelvin, described, e.g., in Refs. \cite{Kel63,Lam82,Cha61,ST83,Gle92}. For quadrupole oscillations in a spherically homogeneous self-gravitating star with radius $r_s$, mass $m_s$, and average density $\rho_0=m_s/(4\pi r_s^3/3)$,  the  assumption of an incompressible fluid yields  a natural oscillation frequency  
\begin{equation}
\omega_0^2 ={{\cal K}_2 \over {\cal M}_2} = {16 \pi G\rho_0\over 15} = {4Gm_s\over 5r_s^3},\label{Kelvin}
\end{equation}
where ${\cal M}_2$ and ${\cal K}_2$ are the respective quadrupole inertia and stiffness parameters.  
For small amplitude quadrupole oscillations the inertia parameter has been deduced in Ref. \cite{Nix} from which one also obtains the stiffness parameters, namely,
\begin{equation}
{\cal M}_2={3m_sr_s^2\over 10}   , \quad\quad {\rm and}  \quad {\cal K}_2={6Gm_s^2\over 25r_s}.\label{Kelvin2}
\end{equation}
The stiffness arises from increase of gravitational energy due to the quadrupole deformation from a spherical star shape.  Stellar oscillation damping is difficult to model as it can arise from  ``gas", ``radiation", and ``turbulence" contributions, each of them varying wildly over temperature, density, and other properties of the stellar interior. For a gas the viscosity varies as  $\gamma_g \sim T^{5/2}$ whereas for radiation $\gamma_r \sim T^4$. In the presence of turbulence, $\gamma_t =R_e \gamma_r/3$, where $R_e$ is the Reynolds number and the viscosity is several orders of magnitude larger than the radiative (or Jeans) viscosity \cite{Kop64}. In the absence of turbulence, radiation damping dominates over gas viscosity. To avoid dealing with specific stellar conditions, we assume a friction coefficient of the form $\gamma = A_\gamma {\cal M}_2 \omega_0$, with $A_\gamma$ taken as a free parameter. 

The velocity distribution of nearby stars ($\lesssim 100$ pc), obtained with the Hypparcos satellite, shows a non-negligible number of stars moving at speeds in excess of 100 km/s \cite{Bov09}.  Hypervelocity stars, with $v \gtrsim 1000$ km/s are rare, and able to escape the galaxy, but have already been observed \cite{Bro05}. To maximize the effect we are looking after we consider an FSO moving at a high speed, $v=1000$ km/s, relative to the star.  For a collision with an impact parameter $b$, the ``collision time", i.e. the time during which the gravitational force is most effective, is $t_{coll} \sim b/v$. For a collision with $b = 5 r_\odot$  and $v=1000$ km/s, one gets
$t_{coll} \sim 1$ h. The period of oscillations associated with Eq. \eqref{Kelvin} for a solar-type star is  $t_{osc} =2\pi/\omega_0  \sim$ 3 h. Hence, we expect a resonating response of oscillations in this system for impact parameters in the range of a few  times $r_s$.

The stellar oscillations can be disentangled into a mixture of transverse and longitudinal oscillations, as displayed in Figure \ref{fig:tosc}. For collisions with impact parameters equal to  $5r_s$ and larger and for velocities $v \sim 1000$ km/s, the orbital eccentricity is large for solar-type stars, and the hyperbolic orbits become nearly straight lines. Therefore, we can safely consider transverse ($t$) and longitudinal ($l$) oscillations as being those transverse and along the asymptotic velocity, respectively.   The equations of motion for small forced harmonic oscillations can be derived from Eq. \eqref{Fvib4} in terms of the inertia and stiffness parameters of Eqs. \eqref{Kelvin} and \eqref{Kelvin2}, yielding
\begin{equation} \label{eq:Newton}
f_i(t) = \ddot{x}_i(t) + \beta \dot{x}_i(t) + \omega_0^2 x_i(t), \quad {i=t,l}, 
\end{equation}
along the two directions, where $f_i(t)$ is the driving tidal force per unit mass, and $\beta = A_\gamma\omega_0$. For a straight line  trajectory ($R^2=b^2 +v^2t^2$) with no coupling among the orthogonal oscillations, this problem is solvable in analytical form. For hyperbolic trajectories with large eccentricities, our simulations show that accurate results can be obtained replacing the impact parameter $b$ by $b'=a$ in the analytical solutions  below, with $a$ equal to the distance of closest approach at the periapsis.  

\begin{figure}[tbp]
\begin{center}
\includegraphics[scale=0.28]{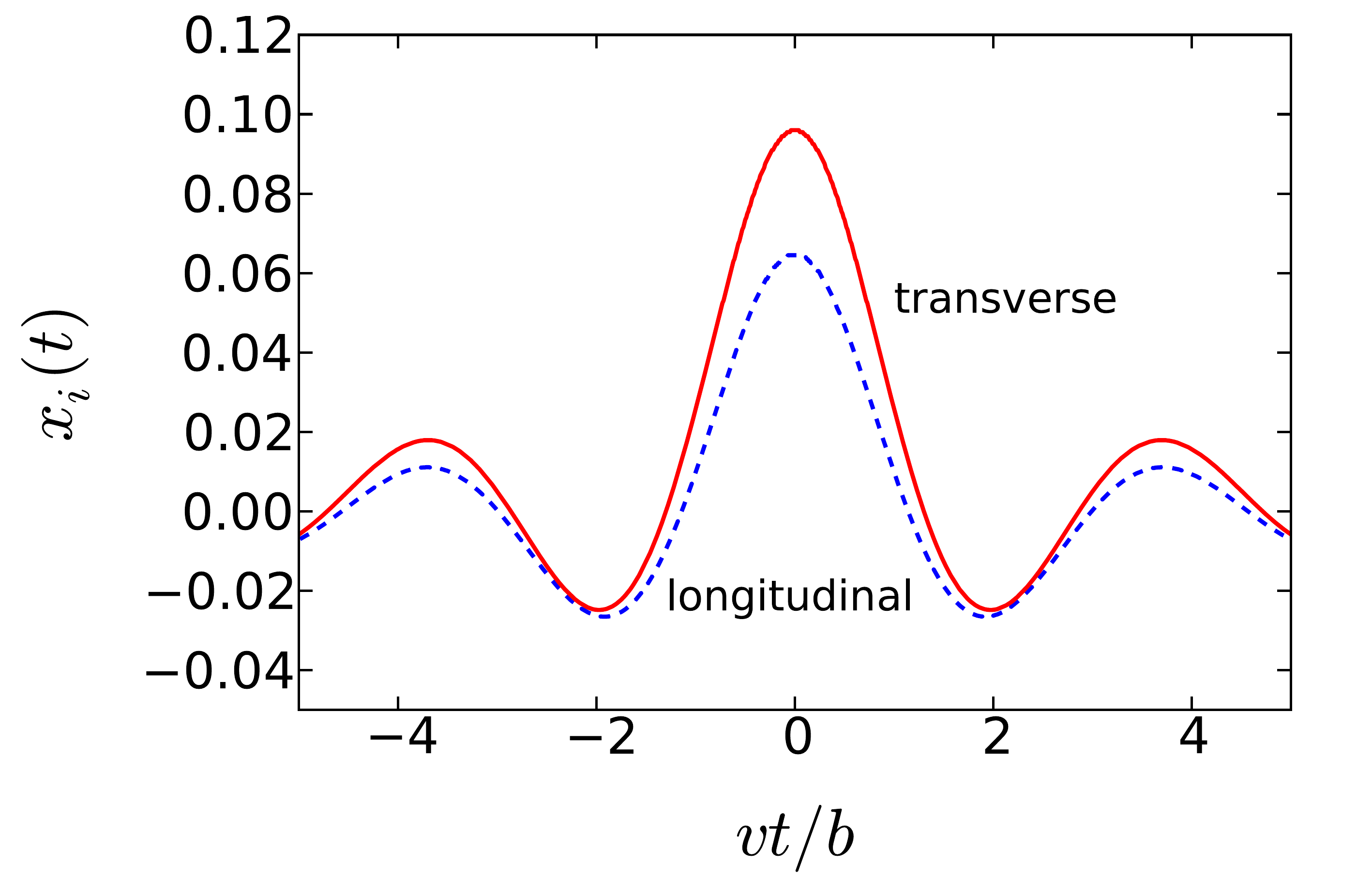}
\caption{Oscillation amplitudes in units of the stellar radius  as a function of time. The same parameters were used as in Figure \ref{fig:xi}.} \label{fig:xit}
\end{center}
\end{figure}

The solution of Eq. \eqref{eq:Newton} is expressed in terms of the Fourier transform $x_i(t)= (2\pi)^{-1/2}\int {\tilde x_i} (\omega) \exp(i\omega t) d\omega$. For a straight line trajectory with effective impact parameter $a$, the amplitudes $\tilde{x}_i(\omega)$ are given by
\begin{equation}
 \tilde{x}_t(\omega)  = \left(\frac{8}{\pi}\right)^{1/2} {GM\over av}  { \xi K_1 \left( \xi \right) \over \left[(\omega_0^2-\omega^2)^2+4  \beta^2 \omega^2\right]^{1/2}},
\end{equation}
where $K_1$ is the first order modified Bessel function, and
\begin{equation}
 \tilde{x}_l(\omega)  = i\left(\frac{8}{\pi}\right)^{1/2} {GM\over av}  { \xi K_0 \left( \xi \right) \over \left[(\omega_0^2-\omega^2)^2+4  \beta^2 \omega^2\right]^{1/2}},
\end{equation}
with $K_0$ the corresponding  zeroth order modified Bessel function. The ``adiabacity" parameter $\xi = \omega a/v$ measures the degree to which the star responds adiabatically to the driving tidal force.  The function $\xi K_1(\xi)$ is nearly constant ({$K_1 \sim 1/\xi$)  for $\xi< 1$, and decays exponentially $K_1 \sim \exp(-\xi)$  for $\xi > 1$.  Hence, oscillation modes with frequencies up to $\omega \sim v/a$ will be preferred and those with larger frequencies will be suppressed exponentially. 

\begin{figure}[t]
\begin{center}
\includegraphics[scale=0.33]{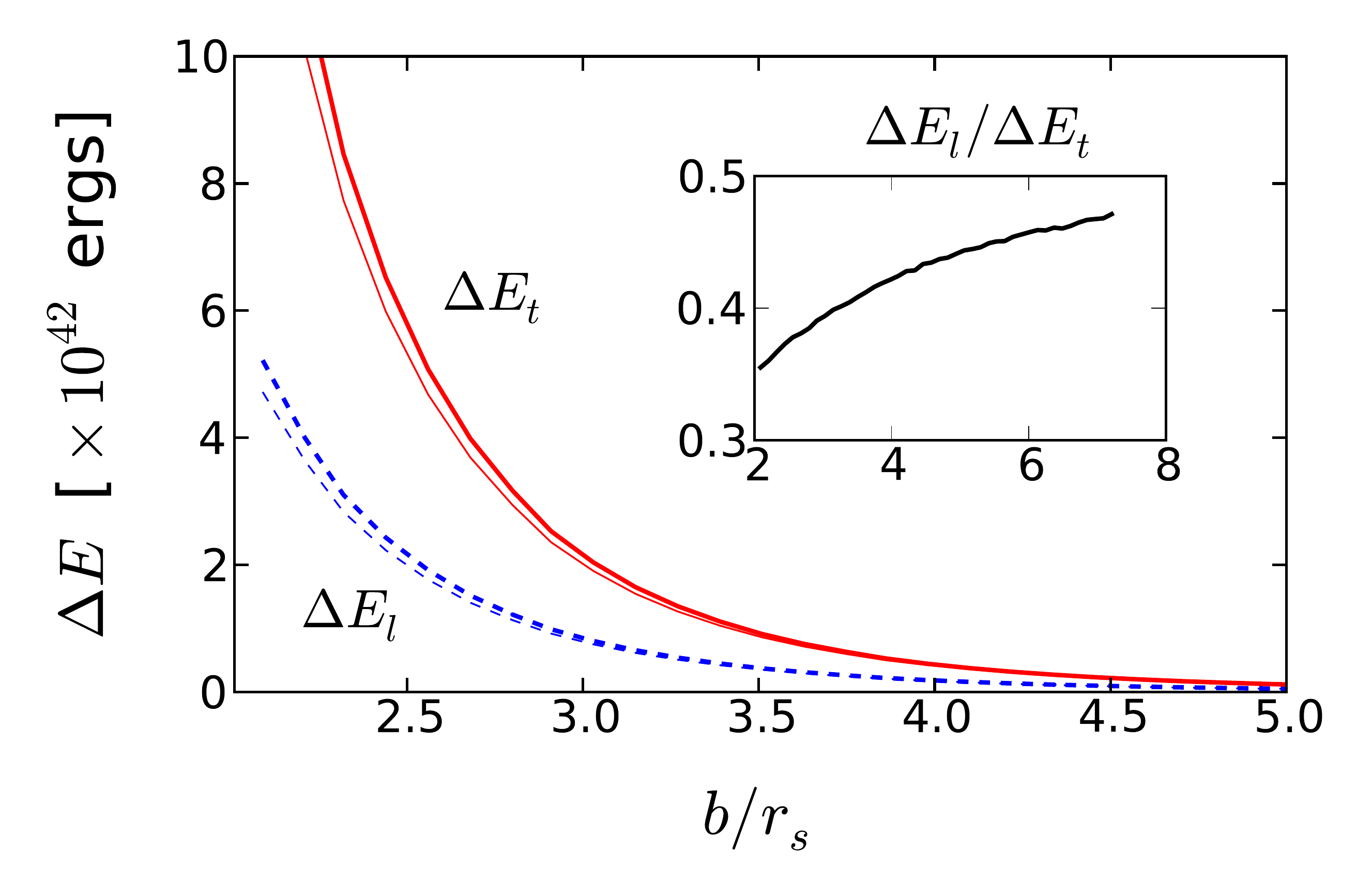}
\caption{Energy, in ergs, transferred to oscillations in a solar-type star along ($l$ - longitudinal) and perpendicular ($t$ - transverse) the direction of incidence of a fast stellar object as a function of the impact parameter and in units of the star radius. The thin lines represent calculations obtained with parametrized hyperbolic trajectories. The inset shows the ratio between the two transferred energies as a function of the same impact parameter measure.} \label{fig:elt}
\end{center}
\end{figure}

In Figure \ref{fig:xi} we plot the amplitudes $|\tilde{x}_i(\omega)|^2$ (multiplied by $10^{-22}$ m$^2$s$^2$) as a function of frequency, for longitudinal (dashed line) and transverse (solid line) oscillations. We used $M=2m_\odot$, $b = 5 r_\odot$,  $v=1000$ km/s, $m_s = m_\odot$, $r_s=r_\odot$, and $A_\gamma=0.1$. Radii and masses are taken in units of the solar mass, $m_\odot$, and radius, $r_\odot$, respectively. We observe a remarkable resonant condition for these choice of parameters. As expected,  the resonance peak decreases as the stellar viscosity increases (increasing  $A_\gamma$). The resonance peak also decreases with the stellar radius  because  the star gets  stiffer as the radius decreases, if its mass is kept constant. In this case the natural oscillation frequency $\omega_0$ becomes large and the resonance matching condition $b/v \sim 1/\omega_0$ does not take place, except for very high FSO velocities, beyond reasonable expectations from present observations. Keeping the same parameters above but varying $r_s$, we conclude that for  white dwarfs (WD) ($r_s \sim r_\odot/10^2$) and neutron stars (NS) ($r_s \sim r_\odot/10^5$) the natural oscillation frequency is too high to match the resonant condition. 

Notice that the collision mechanism discussed here is different than the stellar tidal disruption or breakup in a head-on collision of either a black-hole or a neutron star with another neutron star \cite{LS74}, or those induced in mergers in binary systems \cite{PT77,Tsa12}. A distant collision with a FSO (unless its mass is very large) is unable to yield a tidal disruption of either a WD or a NS, unless maybe for very small impact parameters (see below and also Ref. \cite{Tsa13}).

In Figure \ref{fig:xit} we show the time-dependent  oscillation displacements from equilibrium in units of the stellar radius with the same parameters used in  Figure \ref{fig:xi}. At the periapsis the oscillation amplitudes can reach 10\% of the star radius. This is a large amplitude oscillation, unprecedented by any known observation.  Evidently, for large amplitudes one  expects a non-linear behavior of the oscillations, requiring a more sophisticated model than adopted here. The stellar oscillations start well before the FSO reaches the periapsis  ($t=0$) and are largest at $t=0$. The results displayed in Figure \ref{fig:xit} are close to resonance. An even larger effect would be obtained for a grazing impact parameter, when the stars nearly touch each other at the periapsis. As expected, induced longitudinal oscillations are smaller than transverse ones, but not by much. The difference between oscillations along the two directions increases for conditions off the resonance region. By increasing the star radius  by a factor of 10, resonance conditions can be achieved even for a distant collision, $b=100 r_s$ and larger.

The energy transferred to the stellar oscillations can be obtained from $\Delta E =\sum_{j=l,t} \int_{-\infty}^\infty F_j(t)\dot{x}_j(t)$ or $\Delta E =-2{\cal R}e\sum_j\int_0^\infty i\omega \tilde{x}_j(\omega)\tilde{F}_j^*(\omega) d \omega .$
%\begin{eqnarray}
%\Delta E &=&\sum_{j=l,t} \int_{-\infty}^\infty F_j(t)\dot{x}_j(t)dt\nonumber\\&=&-2{\cal R}e\sum_j\int_0^\infty i\omega \tilde{x}_j(\omega)\tilde{F}_j^*(\omega) d \omega . \label{fig:DE}
%\end{eqnarray} 
The momentum transferred to the recoil, or center of mass motion, of the star is approximately given by $\Delta p_{cm} = 2GMm_s/bv$ and the recoil energy by $\Delta E_{cm}=(\Delta p)^2_{cm}/2m_s$. For a collision with $M=2m_\odot$, $b = 5 r_\odot$,  $v=1000$ km/s, $m_s = m_\odot$, $r_s=r_\odot$,  we get $\Delta E_{cm} = 5.5\times 10^{46}$ ergs,  e.g., 0.3\% of the FSO bombarding energy is transferred to recoil. A much smaller energy is transferred to tidal oscillations. Using  $A_\gamma=0.1$, one obtains  $0.49\times 10^{41}$ ergs and $1.6\times 10^{41}$ ergs transferred to longitudinal and transverse stellar oscillations, respectively. This is larger than energies emitted in X-ray bursts from accretion in binary systems. However, this energy is transferred to the star (and possibly released in form of radiation)  in a much larger time scale: a few hours instead of seconds as in X-ray bursts. Only a fraction $2.8\times 10^{-6}$ of the recoil energy goes into internal excitation of the star. But assuming this star radiates all this energy in form of light with the sun's luminosity ($3.9\times10^{26}$ W), it would be enough for 1.3 years of steady solar luminosity. An appreciable amount of this energy may be emitted in long wavelength radiation   of   long duration, i.e., within few hours. The  characteristics of this radiation depend  on many intrinsic  stellar properties.

Figure \ref{fig:elt} shows the energy in ergs transferred to longitudinal (dashed line) and transverse (solid line) oscillations in a solar mass star as a function of the impact parameter in units of the star radius. We use the same parameters  for $M$, $r_s$, $m_s$ and $A_\gamma$ as in Figure \ref{fig:xi}.  The thin lines show the results obtained with exact hyperbolic trajectories. Only for small impact parameters there is a visible deviation from the results using straight-line trajectories with recoil correction.  The inset shows that the ratio between the two energies increases in the same impact parameter range. For large $b$  the longitudinal contribution becomes as relevant as the transverse one.

%-------------------------------------------------------------
%-------------------------------------------------------------
\begin{table}[ht]
\begin{center}
\begin{tabular}{||c||c|c||c|c||} \hline
$r_s$ (km)   & $\Delta E_{l}$ (ergs) &$\Delta E_{t} $ (ergs) & $x_{max}/r_s$    \\
\hline
\hline
10  (NS)   & $1.09\times 10^{2}$  & $2.09 \times 10^{2}$ &$0.986\times 10^{-22}$ \\
$7 \times 10^3$ (WD)    & $1.76 \times 10^{32}$ & $1.88\times 10^{32} $ &$1.07\times 10^{-5}$ \\
$7 \times 10^5$  (ST) & $0.324\times 10^{42}$ & $0.608 \times 10^{42}$&0.128\\ 
\hline
\end{tabular}
\end{center}
\caption{\label{table:results} The longitudinal and transverse energy transferred to a neutron star (NS), white dwarf (WD) and a solar-type (ST) star, all with masses $m_s=1.4m_\odot$,  due to a collision with a fast stellar object with mass $M=2 m_\odot$ passing by an impact parameter $b=5r_\odot$. The first column lists the assumed radius for the star. The last column gives the maximum tidal displacement in units of the assumed stellar radius.}
\end{table} 
%-------------------------------------------------------------
%-------------------------------------------------------------

The resonant conditions for induced oscillations by a FSO are ideal for solar-type stars. But it is worthwhile to investigate what happens in the case of a neutron star (NS) or a white dwarf (WD). In Table \ref{table:results} we show the longitudinal and transverse energy transferred to a NS, WD and a solar-type star, all with masses $m_s=1.4m_\odot$,  due to a collision with a fast stellar object with mass $M=2 m_\odot$ passing by an impact parameter $b=5r_\odot$. The first column lists the assumed radius for the star. The last column gives the maximum tidal displacement in units of the assumed stellar radius. One observes a dramatic change in the energy transfer due to the smaller star size in contrast to a solar-type star. For neutron stars the energy transfer is negligible. The larger stiffness of a compact star corresponds to a large natural frequency, thus quenching the aforementioned resonant condition. 

The results in Table  \ref{table:results} are for $b=5r_\odot$.  But compact stars also allow closer collisions if the FSO is a WD, a NS, or a BH. Table  \ref{table:results2} shows the same as in Table \ref{table:results} but for closer encounters  of the FSO with a WD and a NS. The collision impact parameter $b$ is measured in units of 5 times the WD (rows 2 and 3) radius, or 5 times the NS radius (row 4). In these cases, the trajectories are significantly modified by the gravitational attraction and we solve Eq. \eqref{Fvib4} parametrized by a hyperbolic trajectory. A close collision of a FSO and a NS might require the solution of general relativity equations for the trajectory, which we do not consider. Our results show that the energy emitted over a  few hours is well below those of known cosmic cataclysmic events, such as gamma-ray bursts \cite{VA09}, but not worthless more investigation.

%-------------------------------------------------------------
%-------------------------------------------------------------
\begin{table}[ht]
\begin{center}
\begin{tabular}{||c||c|c||c|c||} \hline
b&$r_s$ (km)   & $\Delta E_{l}$ (ergs) &$\Delta E_{t} $ (ergs) & $x_{max}/r_s$    \\
\hline
\hline
$5r_{WD}$&10  (NS)   & $4.56\times 10^{27}$  & $1.92 \times 10^{28}$ &$4.17\times 10^{-6}$ \\
$5 r_{WD}$&$7 \times 10^3$ (WD)    & $1.75 \times 10^{43}$ & $3.28\times 10^{43} $ &$0.101$ \\
$5r_{NS}$&10  (NS)   & $1.19\times 10^{46}$  & $2.27 \times 10^{46}$ &$0.102$ \\
\hline
\end{tabular}
\end{center}
\caption{\label{table:results2} Same as in Table \ref{table:results} but for closer encounter collisions of the FSO with an WD and a NS. The collision impact parameter $b$ is measured in units of 5 times the WD (rows 2 and 3) radius, or 5 times the NS radius (row 4). }
\end{table} 
%-------------------------------------------------------------
%-------------------------------------------------------------

In the case of NS-NS collisions, our calculated energy transfer is $\sim 10^{46}$ ergs at $b=5r_{NS}$. Notice that we do not explore the equation of state of nuclear matter, relying solely on the physics of an incompressible fluid. According to Ref. \cite{Tsa12}, this energy would induce high frequency seismic oscillations in the NS which can couple to the magnetic field and spark a particle fireball burst. For solar-like stars and WDs, a close encounter with an ultrafast and ultramassive FSO can lead to {\it stellar fission}, similar to those occurring in a stretched water droplet.  Although rare, such phenomena would be amenable to observation.

We thank beneficial discussions with Seung-Hoon Cha and Kurtis Williams. C.B. and W.N. also acknowledge support under U.S. DOE Grant DDE- FG02- 08ER41533, the NASA Astrophysics Theory Program, Grant 10-ATP10-0095 and the Cottrell College Science Awards.


\begin{thebibliography}{}
\bibitem{Col03}``The Fundamentals of Stellar Astrophysics", George W. Collins, II, NASA Astrophysics Data System (ADS) (2003).
\bibitem{ST83}S.  L. Shapiro and  S.  A.  Teukolsky,    Black   Holes,    ``White  Dwarfs and   Neutron Stars"
(Wiley,  1983).\bibitem{Ter98} C. Terquem, J.C.B. Papaloizou, R.P. Nelson and D.N.C. Lin, Astrophys. J. 502, 788 (1998). 
\bibitem{SW02} G. J. Savonije and M. G. Witte, Astrophys. \& Astron. 386, 211 (2002).
\bibitem{Abb08} B. Abbott et al. (LIGO Scientific Collaboration),  Phys. Rev. D 77, 062002 (2008).
\bibitem{LS74} J.M. Lattimer and D.N. Schramm,  Ap. J. 192, L145 (1974).
\bibitem{Pac86} B. Paczynski, Astrophys. J. Lett. 308, L43 (1986).
\bibitem{Goo86} J. Goodman, Astrophys. J. Lett. 308, L47 (1986).
\bibitem{Rob11} L.F. Roberts, D. Kasen, W.H. Lee, and E. Ramirez-Ruiz,  Astroph. J. Lett. 736, L21 (2011).
\bibitem{MB12} B.D. Metzger and E. Berger, Ap. J. 746, 48, (2012).
\bibitem{Tsa13} D. Tsang,  Astrop. J.  777, 103 (2013).
\bibitem{Avs77} I.N. Avsiuk, Soviet Astronomy Letters, 3, 96 (1977).
\bibitem{Brad08} Hale Bradt, Astrophysics Processes, Cambridge University Press (2008), Chapter 4.
\bibitem{Ja98} J.D. Jackson, ``Classical Electrodynamics",  John Wiley and Sons, Third Edition.
\bibitem{Mou84} ``An Introduction to Celestial Mechanics", F. R. Moulton, Dover Publications (1984).
\bibitem{Dals14} ``Stellar Oscillations", Jorgen Christensen-Dalsgaard, Aarhus Universitet (2014).
\bibitem{Kel63} Sir Thomson W (Lord Kelvin) 1863 Phil. Trans. (papers iii, 384)
\bibitem{Lam82} H. Lamb, London Much. Sot. Proc. 13, 278 (1882).
\bibitem{Cha61} ``Hydrodynamic and  Hydromagnetic Stability", S. Chandrasekhar, Clarendon Press, (1961).
\bibitem{Gle92} N. K. Glendenning, F. Weber and  S. A. Moszkowski, Phys. Rev. C45, 844  (1992).
\bibitem{Nix} J. R. Nix, Ann. Phys. (N. Y.) 41, 52 (1967).
\bibitem{Kop64} Z. Kopal, Astrophysica Norvegica, 25, 239 (1964).
\bibitem{Bov09} Jo Bovy, David W. Hogg, Sam T. Roweis, Astrophys. J. 700, 1794 (2009).
\bibitem{Bro05} W. R. Brown, M. J.  Geller, S. J. Kenyon, and M. J.  Kurtz, Astrophys. J. 622,  L33 (2005).
\bibitem{Tsa12} D. Tsang, J. S. Read, T. Hinderer, A. L. Piro, and R. Bondarescu, Phys. Rev. Lett. 109, 071102 (2012).
\bibitem{PT77} W.H. Press and S.A. Teukolsky, Ap. J. 213, 183 (1977).
\bibitem{FGP09} V Ferrari, L Gualtieri and F Pannarale, Class. Quantum Grav. 26, 125004 (2009).
\bibitem{VA09} G. Vedrenne and J.-L. Atteia,  ``Gamma-Ray Bursts: The brightest explosions in the Universe". Springer/Praxis Books. ISBN 978-3-540-39085-5. (2009).
\bibitem{Tsa12} D. Tsang, J.S. Read, T. Hinderer, A. L. Piro, and R. Bondarescu,  Phys. Rev. Lett. 108, 011102 (2012).


\end{thebibliography}
\end{document}